\documentclass[%
 reprint,
 amsmath,amssymb,
 aps,
 prb,
]{revtex4-1}
\usepackage{mathtools}
\usepackage{amssymb}
\usepackage{graphicx}
\usepackage{amsmath}
\usepackage{amsfonts}
\usepackage{amsthm}
\usepackage{float}
\usepackage{booktabs}
\usepackage{multirow}
\usepackage{pgfplots}
\usepackage{upgreek}
\usepackage{wrapfig}
\usepackage{enumitem}

\usepackage{physics}
\usepackage{tensor}
\usepackage{bbm}
\usepackage{soul}
\usepackage[utf8]{inputenc}
\usepackage[T1]{fontenc}

\usepackage{graphicx}
\usepackage{dcolumn}
\usepackage{bm}

\newcommand{\cL}{\mathcal{L}} 
\newcommand{\cR}{\mathcal{R}} 
\newcommand{\cG}{\mathcal{G}} 
\newcommand{\Ads}{\text{AdS}} 
\newcommand{\beq}{\begin{eqnarray}}
\newcommand{\eeq}{\end{eqnarray}}

\begin{document}
\title{Second-order Lovelock Gravity from Entanglement in Conformal Field Theories}
\author{ Cunwei Fan$^1$, Gabriele La Nave$^2$  and Philip W. Phillips$^1$}
   \affiliation{$^1$Department of Physics and Institute for Condensed Matter Theory,
University of Illinois
1110 W. Green Street, Urbana, IL 61801, U.S.A.}
\affiliation{$^2$Department of Mathematics, University of Illinois, Urbana, Il. 61801}
 
\begin{abstract}
    Holographic entanglement entropy and the first law of thermodynamics are believed to decode the  gravity theory in the bulk.
    In particular, assuming the Ryu-Takayanagi (RT)\cite{ryu-takayanagi} formula holds for ball-shaped regions on the boundary around CFT vacuum states implies\cite{Nonlinear-Faulkner} a bulk gravity theory  equivalent to Einstein gravity through second-order perturbations. In this paper, we show that the same assumptions can also give rise to second-order Lovelock gravity. Specifically, we generalize the procedure in \cite{Nonlinear-Faulkner} to show that the arguments there also hold for Lovelock gravity by proving through second-order perturbation theory, the entropy calculated using the Wald formula\cite{Wald_noether} in Lovelock also obeys an area law (at least up to second order). Since the equations for second-order perturbations of Lovelock gravity are different in general from the second-order perturbation of the Einstein-Hilbert action, our work shows that the holographic area law cannot determine a unique bulk theory even for second-order perturbations assuming only RT on ball-shaped regions.  It is anticipated that RT on all subregions is expected to encode the full non-linear Einstein equations on asymptotically AdS spacetimes.
    
    \end{abstract}
    
    \maketitle
\section{Introduction}

While the AdS/CFT correspondence can be construed broadly as a mechanism for constructing a dual representation of a conformal field theory in terms of gravity, it actually implies an equivalence between the states in both theories.  That is, the quantum states of the conformal theory are actually dual states in the gravitational theory thereby providing a prescription for the representation of states in a quantum theory of gravity. Precisely which conformal theories admit such a description remains  an open question.  What has emerged as a promising route\cite{ ryu-takayanagi,maldacenapert1,van1,RT2} to solving this problem is to show that spacetime is encoded in entanglement\cite{bombelli,Srednicki,heatkernel,callanGeometricEntropy1994,Hawking2001,Calabrese,ch2009,eisert}.  Building on earlier results of Maldacena\cite{maldacenapert1} on a first-order perturbative theory, Faulkner, et al.\cite{Linear-Faulkner}  showed the Hilbert-Einstein action to second order can be deduced starting just from the first-law of thermodynamics and the RT formula on ball-shaped regions. This result is intriguing because it establishes a hard mutual connection between the triumverate in physics: gravity, thermodynamics and quantum mechanics.

We show in this paper that Lovelock\cite{lovelock} gravity can also emerge from entanglement \'a la RT\cite{ryu-takayanagi,RT2}, a fact that seems to indicate that string theoretic corrections to the bulk may go through Lovelock perturbations of vacuum AdS (which has the feature of satisfying both Einstein and Lovelock gravity).  This is important because non-linear curvature theories are common-place in string-theoretic constructions and cosmology.  Consider for example that Lifshitz spacetimes\cite{hkw}   described by the action
\begin{equation}
    S = \frac{1}{16 \pi G_{D}} \int d^{D} x \sqrt{-g} \left ( R -2\Lambda- e^{2\alpha \phi} F^2 - 2|\nabla \phi|^2  \right ), 
\label{dil}
\end{equation}
 admit black brane solutions asymptotic to AdS at infinity, supporting Lifshitz horizons with a scalar dilaton that runs logarithmically. Both electrically and magnetically charged black branes give rise to such geometries. In the former, the dilaton runs towards weak coupling at the horizon ($g\equiv e^{2\alpha \phi} \to 0$), while in the latter, the dilaton runs towards strong coupling.
String theoretic corrections can be analysed by replacing $e^{2\alpha \phi}$ with the {\it gauge coupling function} $f(\phi)$ taking the form
$f(\phi)= \frac{1}{g^2}+ \sum _k \xi _k  g^{2k}$.
The importance of non-linear curvature theories is also manifest in the string theoretic analysis of quantum bulk effects at the near horizon geometries of Lifshitz solutions. 
The effects of these quantum corrections is seen to imply a flow from the UV fixed point (the CFT at infinity) towards the Lifshitz invariant theory where it lingers for a long (and definite) while but then quantum effects take over and smooth the geometry to an $AdS_2 \times \mathbb R^2$ geometry. In \cite{kl} these effects are seen to drive the deep interior region to be replaced by a relativistic fixed geometry such as $AdS_2 \times \mathbb R^{d-1}$ with $D = d +1$.  In fact, they considered  a toy model with coupling $g(\phi)C_{\mu\nu\rho\sigma}^2$ (where $C_{\mu\nu\rho\sigma}$ is the Weyl tensor) and demonstrated that such a term can both stabilize the dilaton and resolve the Lifshitz horizon to $AdS_2 \times \mathbb R^2$.  Further generalizations have also been made based on actions with higher-order curvatures\cite{haehl2018higher}.

\subsection{Key Results}

The central result of this paper is to prove that the construction in\cite{Nonlinear-Faulkner} does not imply only Einstein gravity but Lovelock gravity as well. In particular, we prove that given any CFT theory , and for any state of the form 
\begin{equation}
    \left\langle\varphi_{(0)}| \psi_{\lambda}(\varepsilon)\right\rangle =\int ^{\varphi_{(0)}}e^{-\int_{-\infty}^{0} d x_{E}^{0} d^{d-1} \mathbf{x}\left(\mathcal{L}_{C F T}+\lambda_{\alpha}(x ; \varepsilon) \mathcal{O}_{\alpha}(x)\right)}, 
    \label{eq:perturbation}
\end{equation}
there always exits a metric of the form $g = g_{\Ads} + \epsilon h_1 + \epsilon^2 h_2$, that correctly computes the entanglement entropy up to second order in $\epsilon$ for all ball-shaped regions via the Ryu-Takayanagi formula to \textit{second order}. Here, $\varphi_{(0)}$ is the field configuration on $x_E^0=0$ surface, that is $\varphi_{(0)}= \varphi\left(x_{E}^{0}=0, \mathbf{x}\right)$. Moreover, such metrics must satisfy the second-order Lovelock equation and the stress energy tensor is formed from the matter fields that solve the linearized Lovelock equation about AdS spacetime with boundary conditions specified by the CFT one-point function of $\mathcal{O}_{\alpha}$.  In the Einstein bulk case of \cite{Nonlinear-Faulkner}, one needs to require that $\tilde{C}_T = a^*$. In our case, we can remove such a constraint since, as commented in \cite{Nonlinear-Faulkner}, one can always find appropriate constants in the Lovelock Lagrangian so as to satisfy that condition. 

One of the main observations of this article stems from the fact that we can assume the Ryu-Takayanagi formula for ball-shaped regions just to second order in $\epsilon$. This is a reasonable assumption since we want to determine the equation of motion of the perturbed metric up to second order. Also, we should notice that in general, the holographic entanglement entropy for ball-shaped regions in a CFT  dual to a theory of Lovelock gravity should not follow an area law. However, the results in this paper only show that the entropy is an area law up to second-order perturbations. 
The second related important ingredient is the fact that through $O(\epsilon^2)$, we can calculate the entanglement entropy using the Wald entropy, since it coincides up to second order with the RT formula, at least for perturbations of the vacuum AdS (cf. section \ref{section:ryu-takayanagi_in_lovelock_gravity}\cite{ryu-takayanagi}).
It is not believed\cite{iwashita,deBoer2011,hung2011holographic,chen2013note} that the Wald entropy is the natural holographic entanglement entropy.  Indeed in \cite{dong2014holographic, hung2011holographic}, using a holography with a Lovelock bulk, the CFT entanglement entropy is calculated by the Jacobson-Myers formula. In the discussion section of this paper we show that in Lovelock holography, the Jacobson-Myers formula  for perturbations around vacuum AdS coincides with the RT formula up to second order as well, thus establishing that even the true holographic bulk Lovelock dual to a CFT is indistinguishable from the Einstein bulk dual, up to second order.

Also, we should make clear that we do not claim that the full-fledged bulk theory dual to the boundary CFT should indeed be a Lovelock theory.  In fact, the existence of such a Lovelock bulk theory is a mere feature of the second-order perturbation analysis and information from  higher order perturbations is not determined.  Our bulk Lovelock theory merely captures the holographic entanglement up to second order.  Hence, the key results here support the recent program\cite{parrikar} to encode the full Einstein equations from entanglement on all subregions not just ball-shaped constrictions. The most difficult part of this program is to show the full non-linear equations are satisfied for perturbations of pure AdS. In fact, vacuum AdS has constant curvature, hence it satisfies both Einstein and (all) Lovelock equations of gravity. Intuitively this suggests that AdS lies at the intersection of many branches representing all the moduli spaces of solutions which are all tangent to one another (since first order Einstein and Lovelock coincide) and first order equations cannot suffice to distinguish between all these branches. In this paper we show that not even second order conditions suffice. On the other hand, for {\bf generic metrics} (be it in the Einstein or in one of the Lovelock moduli spaces) we expect that imposing second order conditions should suffice to determine the local perturbations, and thus we expect that for a generic background which is Lovelock (or Einstein), imposing the first law of thermodynamics and the RT formula up to second order, should suffice to determine the bulk field equations. 

Another point we want to make clear is that although Ref. \cite{haehl2018higher} shows that the generalized RT formula can imply a corresponding gravity theory for second order perturbations abound AdS, we show in this paper that the "classical RT", or specifically the area law, would not imply a unique bulk even at second order. This follows because the generalized entanglement formula for Lovelock theory and Einstein theory coincide up to second order perturbations for ball shaped regions, which will be shown in section \ref{section:ryu-takayanagi_in_lovelock_gravity}. Therefore, what Ref.\cite{haehl2018higher} shows is that the entanglement formula corresponding to a specific bulk gravity theory will give the corresponding dynamics of gravity but we showed it is possible in the beginning that two different entanglement formulae of two theories capture the area law up to second order together and thus two bulk theories can be implied from the area law. There is, thus, no obvious conflict between our result and that from Ref. \cite{haehl2018higher}.

\subsection{Summary of derivation}

The derivation of our results follows closely that in \cite{Nonlinear-Faulkner}.
We start with the first law of thermodynamics and the entanglement entropy. Consider, a ball-shaped region $A$ in some CFT theory, and for perturbations in Eq.(\ref{eq:perturbation}), the first law of thermodynamics almost holds  \cite{Nonlinear-Faulkner} and is given by
\beq
    \frac{d}{d \varepsilon}\left(\left\langle H_{A}\right\rangle- S_{A}\right)=\frac{d}{d \varepsilon} S(\rho_{A} \| \rho_{A}^{(0)}),
    \label{eq:relative_entropy}
\eeq
where $\rho_A^{(0)}$ is the density matrix of the region $A$ without perturbations and $S(\rho_A \| \rho_A^{(0})$ is the relative entropy between the perturbed  and unperturbed states. $\expval{H_A}$ is the expectation of the modular Hamiltonian in the entangling region $A$ and its specific form is 
\begin{equation}
    H_A = \int_A d\Sigma^{\mu}T_{\mu\nu} \eta ^{\nu},
\end{equation}
where $\eta^{\nu}$ is a time-like vector that generates the modular flow for region $A$ and $d\Sigma^{\mu}$ is any spacelike surface with boundary as $\partial A$ in the region $\mathcal{D}(A)$ which denotes the union of future and past Cauchy sections.  

\begin{figure}[t]
    \includegraphics[scale=0.3]{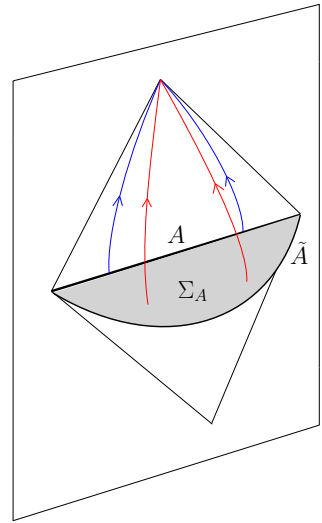}
    \caption{$A$ is a ball-shaped regions on a spatial slice on the boundary. $\tilde{A}$ is the surface that maximizes the Wald functional $S_A^{grav}$ in the bulk. $\Sigma_A$ is the spatial region bounded by $A$ and $\tilde{A}$ on the same slice as $A$. Red arrow lines are the flows of the timelike Killing vector $\xi_A$ and the blue arrow lines are the asymptotic limit of $\xi_A$ which is $\eta^{\nu}$ on the boundary.  }
    \label{fig:rindler_wedge}
\end{figure}

The second main equation needed is the Iyer-Wald equality \cite{Iyer-Wald, Wald_noether}. This equality is an integral form of N\"other's first theorem and it states that the change in the conserved charge on the boundary is related to the change in the currents in the region circumscribed by the boundary. A useful review of this equality and its relation to our problem can be found in \cite{Linear-Faulkner}. 
In what follows, we write down the equality in a convenient form for Lovelock gravity,
\begin{equation}
\begin{aligned}
    \frac{d}{d\epsilon } \left( E_A^{grav} - S_A^{grav}\right) &= \int_{\Sigma_A}\left\{ \omega_{L} (g;h^{(1)}, \cL_{\xi_A} g) \right. \\
    &+ \left. \omega_{\phi}(g ; \frac{d \phi}{d \varepsilon}, \mathcal{L}_{\xi_{A}} \phi) \right\} + \int_{\Sigma_A} \cG .
\end{aligned}
\label{eq:wald_identity}
\end{equation}
Here, $E^{grav}_A$ is the N\"other charge on $A$ and $S_{A}^{grav}$ is the N\"other charge on a bifurcating horizon $\tilde{A}$ which is found by maximizing the functional form of $S_{A}^{grav}$. 
In the literature\cite{iwashita,deBoer2011,hung2011holographic,chen2013note} of Lovelock gravity, $S_{A}^{grav}$ is referred to as the {\it Wald entropy}. Here $\Sigma_A$ is the spacelike region bounded by $A$ and $\xi_A$ is a timelike Killing vector that approaches $\eta^{\nu}$ asymptotically near the asymptotic boundary and vanishes on the bifurcating horizon $\tilde{A}$. This is clearly shown in Figure \ref{fig:rindler_wedge}.  Since $\omega_L$ and $\omega_{\phi}$ are ~2-forms on the phase space of the theory, they are evaluated using pairs of vector fields on the phase space, that is to say pairs of variations of the fields ($g$ or $\phi$).
 Further, $\mathcal{G}$ is a function that depends on the variation $h^{(1)}$ and $d\phi/d\epsilon$ and the Killing field $\xi_A$. It vanishes when the variation $h^{(1)}$ satisfies the linearized Lovelock gravity equation.  As we will show, in section \ref{section:ryu-takayanagi_in_lovelock_gravity}, near pure AdS spacetime (in the perturbation expansion), the Wald entropy for the surface $\tilde{A}$ is proportional to the area of $\tilde{A}$ up to second-order perturbations and as a result, the $\tilde{A}$, which is defined by the surface that minimizes the entropy formula, will be the extremal surface even when the metric is second-order perturbed. 
 This is one of the crucial points in our argument.
 Then, according to \cite{Nonlinear-Faulkner}, we can introduce the Holland-Wald gauge to fix the location of the extremal surface $\tilde{A}$ and to keep $\xi_A$ vanishing on the boundary of $\tilde{A}$.  Thus, when we write formula(\ref{eq:wald_identity}), we implicitly use the Holland-Wald gauge to eliminate the term that would be there due to a variation of embedding of the surface $\tilde{A}$ (in general the HRRT\cite{RT2} surface $\tilde A$ depends on the metric, so if not for the Holland-Wald gauge there would be a term coming from the variation $\tilde A(\epsilon)$) . 

In order to prove our statements, we will firstly assume there exists such a metric perturbed to second order around pure AdS spacetime that calculates the entanglement entropy for spherical regions in the vacuum state of the CFT correctly.  Also, as mentioned above, we will show in section \ref{section:ryu-takayanagi_in_lovelock_gravity} that the Wald entropy for Lovelock gravity for perturbations near pure AdS is proportional to the area for first-order perturbations and if the first-order perturbation is on shell i.e. satisfies the Lovelock equation of motion, the Wald entropy is proportional to area of $\tilde{A}$ for second-order perturbations. However, for now, let us only assume there is an arbitrary first-order perturbation $h^{(1)}$. Since the Wald entropy is proportional to area for first-order perturbation, at $\epsilon = 0$, we can write 
\begin{equation}
    \frac{d}{d\epsilon} S^{grav}_A = C\frac{d}{d\epsilon} \text{Area}(\tilde{A})  + O(\epsilon) = \frac{d}{d\epsilon} S_A + O(\epsilon).
\end{equation}
Here, we used the assumption that the Ryu-Takayanagi formula gives the correct entanglement entropy, that is, $S_A = C A$ for some constant $C$. Furthermore, if we consider (\ref{eq:relative_entropy}) for vanishing radius ball-shaped regions, we will have $\delta S_A = \delta \expval{H_A}$. As a result, we can establish the equality 
\begin{equation}
    \frac{d}{d\epsilon} S^{grav}_A  = \frac{d}{d\epsilon} \expval{H_A} + O(\epsilon) .
    \label{eq:gravity_entropy_cft_energy}
\end{equation}
It is shown in \cite{Linear-Faulkner} that the relation between $T_{\mu\nu}$ and $h^{(1)}_{\mu\nu}$ implied from equality (\ref{eq:gravity_entropy_cft_energy}) gives an extrapolation dictionary. Also, since $E^{grav}_A$ is a function of the asymptotic metric perturbation and $H_A$ is a function of the boundary $T_{\mu\nu}$, we can establish relations between $E_A^{grav}$ with $H_A$ and \cite{Linear-Faulkner} shows they are equal. As a result, we can combine (\ref{eq:relative_entropy}) and (\ref{eq:wald_identity}) to obtain
\begin{equation}
\begin{aligned}
    \frac{d}{d \epsilon} S(\rho_{A} \| \rho_{A}^{(0)}) & = \int_{\Sigma_A}\left\{ \omega_{L} (g;h^{(1)}, \cL_{\xi_A} g) \right. \\
     &+\left. \omega_{\phi}(g ; \frac{d \phi}{d \varepsilon}, \mathcal{L}_{\xi_{A}} \phi) \right\} + \int_{\Sigma_A} \cG + O(\epsilon).
\end{aligned}
\label{eq:punchline_equation}
\end{equation}
We will then do the analysis for first and second-order perturbations to show that the variation of $h_1$ and $h_2$ satisfy the first and second-order Lovelock equations of motion. 

\subsection{First-order}\label{sec:first_order}

The first-order analysis is straightforward since the the relative entropy is known to vanish for first-order perturbations, and the first term on the right-hand side is also zero since $\cL_{\xi_A} g =0 $ for the pure AdS metric and thus, we have 
\begin{equation}
\int_{\Sigma_A} \xi_A^a \delta_1 E_{ab} \epsilon^b = 0
\end{equation}
for all ball-shaped regions, where $\epsilon^b$ is the volume form on $\Sigma_A$. As argued in \cite{Linear-Faulkner}, this means that $\delta_1 E_{ab} =0$. Note, to first-order, there is no stress energy tensor from the matter $\phi$ in the bulk and thus the equation $\delta_1 E_{ab} =0$ is the first-order equation of motion of Lovelock theory in vacuum. 

\subsection{Second order}

From the above argument, we see that the first-order perturbation, $h_1$, satisfies the Lovelock equation of motion near the AdS vacuum,first-order and it will be shown in section \ref{entropy} that this guarantees the Wald entropy $S_A$ is proportional to area for second-order perturbations and therefore we have 
\begin{equation}
\begin{aligned}
    \frac{d}{d\epsilon} S_{A}^{grav}  &= \frac{d}{d\epsilon} \text{Area}(\tilde{A}) + O(\epsilon^2) = \frac{d}{d\epsilon} S_A + O(\epsilon^2) \\
    &= \frac{d}{d\epsilon} \expval{H_A} + O(\epsilon^2).
\end{aligned}
\end{equation}
This implies that $E_A^{grav} = \expval{H_A} $ for second-order perturbation 
and thus we can improve (\ref{eq:punchline_equation}) for second-order perturbations  which means the error in (\ref{eq:punchline_equation}) increases to second-order and as as result, when we take the derivative of (\ref{eq:punchline_equation}), we obtain
\begin{equation}
\begin{aligned}
    \frac{d^{2}}{d \epsilon^{2}} S\left.\left(\rho_{A} \| \rho_{A}^{(0)}\right)\right|_{\epsilon=0}= &\int_{\Sigma_{A}} \omega\left(g^{(0)}, h^{(1)}, \mathcal{L}_{\xi_{A}}  h^{(1)}\right) \\
    &-\int_{\Sigma_{A}} 2 \xi_{A}^{a} E_{a b}^{(2)} \epsilon^{b} + O(\epsilon),
\end{aligned}
\label{eq:second_order_punchline}
\end{equation}
where we denote both $\phi$ and $g$ as $g$. If we can show the left-hand side equals the first term on the right-hand side, this implies the last term on the right-hand side is zero which means the second-order Lovelock gravity equation holds. In general, this involves hard calculations in CFT and the gravity theory. However, it is known\cite{Nonlinear-Faulkner} for Einstein theory that $d^2S/d\epsilon^2  = \tilde{C}_T/{a^*} \omega_{E}(h^{(1)})$. This equality holds if $h^{(1)}$ satisfies the first-order Einstein theory. We will show in section \ref{subsection:lovelock_gravity_compared_to_einstein_gravity}  that the Lovelock form $\omega_L$ is proportional to $\omega_E$ for perturbations when the 2-form $\omega$ is evaluated on the AdS metric. Also, we will show that first-order Lovelock is equivalent to first-order Einstein. Besides, $d^2S/d\epsilon^2  $ only depends on the perturbed CFT states instead of the gravity theory. Thus, for first-order perturbation satisfying Einstein or Lovelock, we can show $d^2S/d\epsilon^2  \sim \omega_{E}(h^{(1)}) \sim \omega_L(h^{(1)})$. Furthermore, we can give constraints for Lovelock theory to make the constant of proportionality unity and as a result, the constraint $\tilde{C}_T = a^*$ is removed. Thus, the equality between $d^2S/d\epsilon^2 $ and $\omega_L(h^{(1)}$ is established and thus the last term in (\ref{eq:second_order_punchline}) vanishes which means the Lovelock equation of motion holds. 
Thus, we will have shown that  if there exists a metric $g = g_{\Ads} + \epsilon h^{(1)} + \epsilon^2 h^{(2)}$ that captures the entanglement entropy via Ryu-Takayanagi formula, then the perturbations must satisfy second-order Lovelock gravity. Since, we can always find $h^{(1)} +  \epsilon h^{(2)}$ that satisfy the second-order Lovelock gravity, such a metric must exist. 

\vspace{1em}

Note in our proof, we did not claim that the Wald entropy is the holographic entanglement entropy. Actually, as shown in \cite{dong2014holographic, hung2011holographic}, for a CFT dual to a Lovelock gravity bulk theory, the holographic entanglement entropy should be calculated via the Jacobson-Myers formula. In the discussion section, we will show that for the Lovelock dual, if we use the Jacobson-Myers formula to calculate the entanglement entropy for perturbations around AdS metric, the entropy will also be proportional to the area of the minimal surface up to second order perturbations. This in turn confirms our result that the Ryu-Takayanagi formula could not determine the bulk theory since there exists theories other than Einstein that captures the entanglement entropy for second-order perturbations via Ryu-Takayanagi.

\section{Lovelock gravity}
\subsection{Lovelock action and AdS solution}

To proceed, we first introduce the Lovelock action.
Lovelock gravity theory is a generalization of Einstein's theory that involves up to second-order derivatives of the metric tensor but is nonlinear in the Riemann tensor. These two requirements imply terms such as $R_{abcd}R^{abcd}$ cannot appear alone since the resultant equations of motion of this term will involve derivatives of $R_{abcd}$ which contain higher-order derivatives of the metric. In this case, after some calculation, the form of the theory which contains $R_{abcd}R^{abcd}$ must also contain two other terms and the theory becomes Gauss-Bonnet. Analogously, for theories with arbitrary powers of $R_{abcd}$, the Lagrangian must be of the form
\begin{equation}
    \cL_{(m)} = \sqrt{-g}L_{(m)}, 
\end{equation}
with 
\begin{equation*}
L_{(m)} = \frac{1}{2^m}\delta^{a_1 b_1 \dots a_m b_m}_{c_1 d_1 \dots c_m d_m} \tensor{R}{^{c_1 d_1}_{a_1 b_1}}  \dots \tensor{R}{^{c_m d_m}_{a_m b_m}},
\end{equation*}
where $m$ is the number of powers in $R_{abcd}$ and $\delta^{a_1 b_1 a_2 b_2\dots a_m b_m}_{c_1 d_1 c_2 d_2\dots c_m d_m} $ is the generalized Kronecker delta. Notice we include the proper volume factor in the definition of the Lagrangian density. The full theory of Lovelock gravity is a theory with linear combinations of those terms. Also, there must exist a constant which depends on the dimension of the bulk, $D$, such that powers of $R$ higher than the constant vanishes. We call this number $M$. We will show that $M$ is the largest integer that is less or equal to $(D-1)/2$. Thus, the full Lagrangian is 
\begin{equation}
    \cL = \sum_{m=0}^M c_m \cL_{(m)} = \sqrt{-g}\sum_{m=0}^M c_m L_{(m)}.
\end{equation}
It is customary to define tensor $P^{abcd}$ and the generalized Ricci tensor $\cR^{ab}$ as 
\begin{equation}
    P^{abcd} = \frac{\delta \cL}{ \delta R_{abcd}} \text{,\quad and \quad } \cR^{ab} = P^{acde} \tensor{R}{^{b}_{cde}}.
\end{equation}
With those definitions, it is easy to check that $\nabla_a P^{abcd} =0$ and therefore, the variation of the action will be 
\begin{equation}
\begin{aligned}
    \delta S = \int d^D x \sqrt{-g}&\left\{ \left(\cR^{ab} - \frac{1}{2}g^{ab}\cL\right)\delta g_{ab}  \right. \\
    &\qquad + \nabla_c \left(2 P^{abcd}\nabla_b \delta g_{ad} \right) \bigg\}.
\end{aligned}
    \label{eq:variation_of_lovelock_action}
\end{equation}
The last term is a boundary term and should vanish if we put the boundary at infinity and impose a Dirichlet boundary condition for $g_{ab}$. The first term should vanish for arbitrary variations of the metric to satisfy the maximization condition and thus, the term in the first bracket is the equation of motion in the absence of matter. However, if we have matter fields that are minimally coupled to the metric and stress energy tensor for the field $T_{ab}$, the equation of motion is 
\begin{equation}
    \cR^{ab} - \frac{1}{2}g^{ab}\cL = 8\pi G T_{ab}.
\end{equation}
From the definition of the $L_{(m)}$ and properties of the Kroneker delta, calculations (as done in \cite{lovelock}) can be done to show that 
\begin{equation}
    \cR^{ab} - \frac{1}{2}g^{ab}L = \sum_m c_m E^{(m)}_{ab},
     \label{eq:lovelock_eqm}
\end{equation}
with 
\begin{equation*}
{E}^{a(m)}_{\; b} = -\frac{1}{2^{m+1}}\delta^{ec_1d_1\dots c_m d_m}_{ba_1b_1\dots a_m b_m} \tensor{R}{^{a_1b_1}_{c_1d_1}} \cdots \tensor{R}{^{a_mb_m}_{c_md_m}}.
\end{equation*}
It should be clear now why the maximal value of power of $R$ is of the form chosen. The Kronecker delta is zero when the number of indices is larger than the dimension and thus when $2m+1 >D$, the $L_{(m)}$ terms vanish from the equations of motion, thereby leading to the advertised value of $M$.

In this paper, we are considering perturbations of the metric around vacuum $\Ads$ and to be expedient, we must check that Lovelock gravity admits such a solution. Indeed, since Ads is a maximally symmetric spacetime, the Riemann curvature tensor of AdS with radius $\alpha$ must be written in the form
\begin{equation}
    R_{abcd} = -\frac{1}{\alpha^2}\left(g_{ac}g_{bd} - g_{ad}g_{bc} \right).
    \label{eq:ads_curvature}
\end{equation}
We can plug in the form of the Riemann curvature tensor into the equation of motion and then check whether the solution satisfies the equations of motion. If we take the $\Ads$  Riemann curvature tensor, Eq.(\ref{eq:lovelock_eqm}), we find 
\begin{equation}
    {E}^{a(m)}_{\; b}= -\left(-\frac{1}{\alpha^2} \right)^m \frac{(D-1)!}{(D-2m-1)!}\delta^a_b,
\end{equation}
and the equation of motion of  Lovelock gravity with cosmological constant $\Lambda$ imply that the cosmological constant is 
\begin{equation}
\begin{aligned}
    \Lambda \delta^a_b &= -2 \sum_{m=1}^M c_m {E}^{a(m)}_{\; b} \\
    &= 2c_m\sum_{m=1}^M  \frac{(D-1)!}{(D-2m-1)!}\left(-\frac{1}{\alpha^2} \right)^m \delta^a_b.
\end{aligned}
\label{eq:cosmological_constant}
\end{equation}
Since the cosmological constant $\Lambda$ is just the coefficient of $L_{(m)}$,  $c_0$, we can always set the value to that above so that a pure AdS spacetime solution exists. 

\subsection{Lovelock and Einstein gravity}

\label{subsection:lovelock_gravity_compared_to_einstein_gravity}
In order to show that the Ryu-Takayanagi formula holds to second order also implies Lovelock gravity to second order, we need to compare Lovelock and Einstein gravity to second order. Here, the perturbation is performed around the pure AdS spacetime and therefore, we can write the metric as 
\begin{equation}
    g_{ab} = g^{\Ads}_{ab} + \epsilon h^{(1)}_{ab} + \epsilon^2 h^{(2)}_{ab},
\end{equation}
and $\epsilon$ is a small parameter controlling the perturbations. The Riemann tensor will also be modified due to the perturbation of the metric and if we collect terms according to the power of $\epsilon$, the Riemann tensor will be 
\begin{equation}
    \tensor{R}{^{ab}_{cd}} = {R}^{ab(0)}_{\quad cd} + \epsilon {R}^{ab(1)}_{\quad cd} + \epsilon^2 {R}^{ab(2)}_{\quad cd}, 
\end{equation}
where ${R}^{ab(0)}_{\quad cd}$ is the Riemann tensor for the unperturbed metric $g^{\Ads}_{ab}$ and ${R}^{ab(1)}_{\quad cd}$ only contains $g_{\Ads}$ and $h^{(1)}$, while ${R}^{ab(2)}_{\quad cd} $ contains  $g_{\Ads}$, $h^{(1)}$ and $h^{(2)}$. 

We can calculate now the equations of motion for the perturbed metric using Eq.~(\ref{eq:lovelock_eqm}) and expand in $\epsilon$. Since each factor of $\tensor{R}{^{ab}_{cd}}$ contains first and second-order perturbations, to first-order perturbation in $\tensor*{E}{^{a(m)}_{\ b}}$, we have $m$ ways to perturb the tensor by replacing one of the $\tensor{R}{^{ab}_{cd}}$ factors by ${R}^{ab(1)}_{\quad cd}$. To second-order perturbations, we have two major ways to perturb the tensor $\tensor*{E}{^{a(m)}_{\ b}}$, one is replacing $\tensor{R}{^{ab}_{cd}}$ with ${R}^{ab(2)}_{\quad cd}$ and the other is replacing two factors of$\tensor{R}{^{ab}_{cd}}$ with two factors of  ${R}^{ab(1)}_{\quad cd}$. Thus, we can write 
\begin{equation}
    \tensor*{E}{^{a(m)}_{\ b}} = \tensor*{E}{^{a(m)}_{\ b}} + \epsilon \delta_1 \tensor*{E}{^{a(m)}_{\ b}} + \epsilon^2 (\delta_2 \tensor*{E}{^{a(m)}_{\ b}} + \delta_3 \tensor*{E}{^{a(m)}_{\ b}}).
\end{equation}
\begin{widetext}
\subsection{First-order equation of motion}
We first focus on the first-order perturbation and using the definition in Eq.~(\ref{eq:lovelock_eqm}), we obtain that
\beq
    \delta_1 \tensor*{E}{^{a(m)}_{\ b}}& = -\frac{m}{2^{m+1}} \delta^{ac_1d_1c_2d_2\dots c_md_m}_{ba_1b_1a_2b_2\dots a_mb_m} {R}^{a_1b_1(1)}_{\quad c_1 d_1} {R}^{a_1b_1(0)}_{\quad c_1 d_1} \cdots {R}^{a_1b_1(0)}_{\quad c_1 d_1} \\
    &= -\frac{m}{4} \left( -\frac{1}{\alpha^2} \right)^{m-1} \frac{(D-3)!}{(D-2m-1)!} \delta^{ac_1d_1}_{bc_1d_1} {R}^{a_1b_1(1)}_{\quad c_1 d_1}\\
    &= m\left( -\frac{1}{\alpha^2} \right)^{m-1} \frac{(D-3)!}{(D-2m-1)!} \left({R}^{a(1)}_{\; b} - \frac{1}{2}\tensor{\delta}{^a_b} R^{(1)}  \right),
\eeq
where ${R}^{a(1)}_{\; b}$ and $R^{(1)}$ are first-order perturbations of the Ricci tensor and Ricci scalar, respectively. 
Obviously, the first-order perturbation to the tensor $\tensor*{E}{^{a(m)}_{\ b}}$ is proportional to linearized Einstein gravity. Since, the variation of the cosmological term $\Lambda \tensor{\delta}{^a_b}$ vanishes, $\delta_1 \tensor*{E}{^{a(m)}_{\ b}}$ vanishes for all orders of $m \geq 1$ and thus, we have 
\begin{equation}\label{eq:linearEinstein}
    {R}^{a(1)}_{\; b} - \frac{1}{2}\tensor{\delta}{^a_b} R^{(1)}  =0.
\end{equation}
By definition, $R^{(1)}_{ab} = h_{ac}R^{c(0)}_{\;b} + g^{\Ads}_{ac}R^{c(1)}_{\;b}   $, we can then, find another form in the equations of motion for the small perturbations, $h_{ab}$ as 
\begin{equation}
    R^{(1)}_{ab} - \frac{1}{2}g^{\Ads}_{ab}R^{(1)} - \frac{1}{2}R^{(0)} h_{ab} =\left(R^{c(0)}_{\; b} - \frac{1}{2} R^{(0)} \delta^{c}_{\; b}\right) h_{ac} + g_{ac}^{AdS} \left( R^{c(1)}_{\; b} - \frac{1}{2} R^{(1)} \delta^{c}_{\; b}\right). 
    \label{eq:first_order_einstein}
\end{equation}
The second term vanishes due to the equation of motion for first-order perturbation. The first term gives a constant if $g_{AdS}$ is a pure AdS solution to Lovelock gravity and the constant is 
\begin{equation}
    R^{(1)}_{ab} - \frac{1}{2}g^{\Ads}_{ab}R^{(1)} - \frac{1}{2}R^{(0)} h_{ab} = \left(R^{c(0)}_{\; b} - \frac{1}{2} R^{(0)} \delta^{c}_{\; b}\right) h_{ac} = \frac{(D-1)(D-2)}{2\alpha^2} h_{ab}.
\end{equation}
Notice, this is the linearized Einstein equation but with a different cosmological constant than the one in the Lovelock theory. However, this new cosmological constant is the one we expect in pure Einstein theory for  AdS spacetime. Therefore, the first-order Lovelock equation of motion is equivalent to first-order Einstein. 

\subsection{Second-order equation of motion}

Next we calculate the second-order perturbation of $\tensor*{E}{^{a(m)}_{\ b}}$. It should be clear that if we just perturb the Riemann tensor to second order, we will obtain the same form as the first-order perturbation; that is we have 
\begin{equation}
    \delta_2 \tensor*{E}{^{a(m)}_{\ b}} = - \frac{m}{4} \left( -\frac{1}{\alpha^2} \right)^{m-1} \frac{(D-3)!}{(D-2m-1)!} \delta^{ac_1d_1}_{bc_1d_1} \delta_2\tensor{R}{^{a_1b_1}_{c_1 d_1}}
\end{equation}
which is exactly the Einstein equation perturbed to second order up to a constant factor. However, for $m\geq 2$, we can find the second-order perturbation by perturbing two of the Riemann tensors in Eq.~(\ref{eq:lovelock_eqm}) to  first order. Therefore, we write
\begin{equation}
    \delta_3 E^{a(m)}_{\;b} = -\frac{m(m-1)}{16} \left( -\frac{1}{\alpha^2} \right)^{m-2} \frac{(D-5)!}{(D-2m-1)!} \delta^{ac_1d_1c_2d_2}_{ba_1b_1a_2b_2} {R}{^{a_1b_1(1)}_{\quad c_1 d_1}} {R}^{a_2b_2(1)}_{\quad c_2 d_2}.
\end{equation}
This equation shows that the term $\delta_3 E^{a(m)}_{\;b}  $ is a Gauss-Bonnet term for a first-order variation of $\tensor{R}{^{ab}_{cd}}$. If we assume the first-order perturbation satisfies the first-order Lovelock gravity and thus the first-order Einstein, we can simplify the equation to
\begin{equation}
    \delta_3 E^{a(m)}_{\;b} = C_m \left( 4R^{ae(1)}_{\quad cd} R^{cd(1)}_{\quad be} - \tensor{\delta}{^a_b} R^{cd(1)}_{\quad ef} R^{ef(1)}_{\quad cd} \right).
\end{equation}
This term in general does not vanish and therefore, we have shown that the second-order Lovelock and the second-order Einstein equations differ. 

\subsection{Calculation of $\omega_L$}
In order to carry out the argument of our proof, we need to show that the 2-form $\omega_L$ is proportional to the Einstein 2-form $\omega_{grav}$ in \cite{Nonlinear-Faulkner}. 
The definition of the $\omega_L$ is 
\begin{equation}
\omega_L(g;h_1,h_2) = h_1 \frac{\delta}{\delta g} \theta(g;h_2) - h_2 \frac{\delta}{\delta g} \theta(g;h_2),
\end{equation}
where $\theta$ is defined through $\theta(g;h) = 2P^{abcd} \nabla_b h_{ad} \boldsymbol{\epsilon}_a$ (this should really be $P^{abcdef}$ which are the components of the tensor obtained from $P$ and $g$ via $P\mathbin{\bigcirc\mspace{-15mu}\wedge\mspace{3mu}} g$, the Kulkarni-Nomizu product of $P$ and $g$ (Strictly speaking, this is a generalization of the Kulakrni-Nomuzi product, due to Kulkarni and uses the language of double forms), but we are contracting using $P\mathbin{\bigcirc\mspace{-15mu}\wedge\mspace{3mu}} g$ and that's the same as contracting with $P$ first and then taking the Kulkarni product. 

Before we perform our calculations, let us provide a theoretical description of the proof. 
Using this we now claim that calculating $h_1 \delta g \frac{\delta R_{ab cd} }{\delta g}h_2 $ for any two $h_i$ and $h_2$ quadratic forms with at least one of them in the tangent space to the moduli space of Lovelock equation at pure ADS (i.e. it satisfies the linearized Lovelock equation which is the same as the linearized Einstein eq. \eqref{eq:linearEinstein}) is the same, up to a constant of proportionality, to 
calculating $h_1 P_1 h_2$. 
The reason why we restrict ourselves to the tangent space of Lovelock (which is the same as the tangent space of Einstein) is that setting $\epsilon =0$ in eq. \eqref{eq:punchline_equation} one sees that the first order perturbation must satisfy first order Einstein (as we saw in section \ref{sec:first_order})
This can be seen by first observing that from the tensorial properties of the expression $h_1 \delta g \frac{\delta R_{ab cd} }{\delta g}h_2 $, it can be calculated in any coordinate system and we are thus free to choose one such coordinate system centered at an arbitrary point $p$ in which $h_1$, at the point, is  proportional to $\delta _{ab}$ and in which $\delta g _{ab} = \delta _{ab}$ (we can diagonalize two quadratic forms simultaneously at a point). With such choice, using the fact that the linearized Lovelock equations are the same as the Einstein equations (and that at the point $p$ with the coordinate chosen we have that ${h_1}_{ab}\left( \nabla _a \nabla _b \delta g _{cd} - \nabla _a \nabla _c\delta g _{bd}-\nabla _b \nabla _d \delta g _{b}+\nabla _b \nabla _c \delta g _{ad}\right)$ is proportional to $\delta g$ by the first order Lovelock equations (which are the same as Einstein) and we can conclude. So the crucial ingredient of the proof is that pure AdS sits in all the moduli spaces of Einstein and Lovelock equations and that these moduli spaces are all tangent at pure AdS.

We will perform the computation for general $h_1$ and $h_2$ as far as we can and then we will specialize to the case at hand, namely $h_1= h^{(1)}$ and $h_2= L_{\xi _A} h^{(1)}$.  In order to perform the calculations, we first rewrite this in the form 
%
observe that by the Leibniz rule we have
\begin{equation}\label{dtheta}
\frac{\delta}{\delta g} \theta(g;h) = 2\frac{\delta P^{abcd} }{\delta g} \nabla_b h_{ad} \boldsymbol{\epsilon}_a+ 2P^{abcd}\frac{\delta \nabla_b}{\delta g} h_{ad} \boldsymbol{\epsilon}_a.
\end{equation}
\noindent
In order to calculate $\frac{\delta P^{abcd} }{\delta g}$ we first recall the standard explicit form of $P^{abcd} :=\frac{\partial L_{(m)}}{ \partial R_{abcd}} $, in its $(2,2)$-tensor incarnation.

\begin{equation}\label{eqP}
   { P_{(m)}}^{\mu \nu}_{\alpha \beta} =\frac{m}{2^m} \delta ^{\mu \nu \sigma _1 \cdot \sigma _{2m-2}}_{\alpha \beta \lambda _1\cdots \lambda _{2m-2}} R_{\sigma _1\sigma _2}^{\lambda _1\lambda _2} \cdot R_{\sigma _i\sigma _{i+1}}^{\lambda _i\lambda _{i+1}} \cdots R_{\sigma _{2m-3}\sigma _{2m-2}}^{\lambda _{2m-3}\lambda _{2m-2}} \end{equation}  
The reason why we may switch to the $(2,2)$ representation of the tensor is that the piece that we care about, namely  $\frac{\delta P^{abcd} }{\delta g} \nabla_b h_{ad} \boldsymbol{\epsilon}_a$, can also be rewritten as
\begin{equation}
\frac{\delta P^{abcd} }{\delta g} \nabla_b h_{ad} \boldsymbol{\epsilon}_a= \frac{\delta P^{ad}_{bc} }{\delta g} \nabla_b h^{ad} \boldsymbol{\epsilon}_a.
\end{equation}
 We remark that a quick calculation using formula \eqref{eqP} shows that for AdS, one has  
\begin{equation}
    P_{(m)}^{abcd} \mid_{g=g_{AdS}}= \frac{\partial L_{(m)}}{ \partial R_{abcd}} = \frac{m}{4} \left(-\frac{1}{\alpha^2}\right)^{m-1} \frac{(D-2)!}{(D-2m)!}\left( g^{ac}g^{bd} - g^{ad}g^{bc} \right) = m \frac{(D-2)!}{(D-2m)!} \left(-\frac{1}{\alpha^2}\right)^{m-1} P^{abcd}_{(1)} \mid_{g=g_{AdS}}.
\end{equation}
and therefore the second addendum in Eq. \eqref{dtheta} poses no problem.
Next we calculate the variation of the tensor $P$ w.r.t. $g$, using the Leibniz rule
\begin{equation}
\frac{\delta {P_{(m)}}^{\mu \nu}_{\alpha \beta}  }{\delta g} = \frac{m}{2^m} \delta ^{\mu \nu \sigma _1 \cdot \sigma _{2m-2}}_{\alpha \beta \lambda _1\cdots \lambda _{2m-2}} 
\sum _{j=1}^{2m-2} \prod _{i\neq j} R_{\sigma _i\sigma _{i+1}}^{\lambda _i\lambda _{i+1}}
\frac{\delta R_{\sigma _j\sigma _{j+1}}^{\lambda _j\lambda _{j+1}} }{\delta g}
\end{equation}
whence, using that for AdS $R_{\sigma _i\sigma _{i+1}}^{\lambda _i\lambda _{i+1}}=-\frac{1}{\alpha ^2} \delta _{\sigma _i\sigma _{i+1}}^{\lambda _i\lambda _{i+1}}$
\begin{equation}
\frac{\delta {P_{(m)}}^{\mu \nu}_{\alpha \beta}  }{\delta g} \mid_{g=g_{AdS}}=\left( -\frac{1}{\alpha ^2} \right)^{2m-2}\frac{m}{2^m} \delta ^{\mu \nu \sigma _1 \cdot \sigma _{2m-2}}_{\alpha \beta \lambda _1\cdots \lambda _{2m-2}} 
\sum _{j=1}^{2m-2} \prod _{i\neq j} \delta_{\sigma _i\sigma _{i+1}}^{\lambda _i\lambda _{i+1}}
\left(\frac{\delta R_{\sigma _j\sigma _{j+1}}^{\lambda _j\lambda _{j+1}} }{\delta g}\right)\mid_{g=g_{AdS}}
\end{equation}
So far we have shown that $\omega _L$ is proportional to $\omega _{grav}$ provided that  $h^2 \frac{\delta R^{ad}_{bc} }{\delta g} \nabla_b h^{ad} \boldsymbol{\epsilon}_a$ (which is proportional to $h^2 \frac{\delta P^{ad}_{bc} }{\delta g} \nabla_b h^{ad} \boldsymbol{\epsilon}_a$) is proportional to $ h^2 \frac{\delta {P_1}^{ad}_{bc} }{\delta g} \nabla_b h^{ad} \boldsymbol{\epsilon}_a$. We thus have to calculate $\left(\frac{\delta R_{\sigma _j\sigma _{j+1}}^{\lambda _j\lambda _{j+1}} }{\delta g}\right)\mid_{g=g_{AdS}}$ for a perturbation of $g_{AdS}$.
It is a standard fact that 
\begin{equation}\label{eq:var-Riem}
\frac{\delta R_{ab cd} }{\delta g}=-\frac{1}{2} \left( \nabla _a \nabla _b \delta g _{cd} - \nabla _a \nabla _c\delta g _{bd}-\nabla _b \nabla _d \delta g _{b}+\nabla _b \nabla _c \delta g _{ad}\right),
\end{equation}
and therefore for $h^{(1)}=\delta g$, using that it satisfies the Einstein equation as shown in Eq. \eqref{eq:linearEinstein}
\begin{equation}
h^{(1)\,cd} \frac{\delta R_{ab cd} }{\delta g}= R_{ab}^{(1)} = \frac{R^{(1)}}{2} h_{ab}.
\end{equation}
In this equation we have chosen coordinates (at any given point) for which the Christoffel symbols are $0$ at the point (not its derivatives) so that the covariant derivatives in equation \eqref{eq:var-Riem} can be replaced by {\it normal} derivatives. This takes care of the term $h^{(1)\, cd }  \frac{\delta R_{ab cd} }{\delta g} \nabla _b \left( L_{\xi _A}h^{(1)}_{ad}\right)$. The other term, $L_{\xi _A}h^{(1)\, cd }  \frac{\delta R_{ab cd} }{\delta g} \nabla _b h^{(1)}_{ad}$, is a purely gauge contribution and in fact, making use of tensoriality, we can choose coordinates for which $ \nabla _b h^{(1)}_{ad} (p) =0$ at any point $p$.

 Therefore, we have shown that  
\begin{equation}
    \omega_L = \sum_{m=1}^{M} c_m m \frac{(D-2)!}{(D-2m)!} \left(-\frac{1}{\alpha^2}\right)^{m-1} \omega_{grav} = B \omega_{grav} .
\end{equation}
Consequently, from the derivation in Ref. \cite{Nonlinear-Faulkner}, the above results imply that
\begin{equation}
    \delta^{(2)} S\left(\rho_{A}| | \rho_{A}^{(0)}\right)= \frac{1}{B}\frac{\widetilde{C}_{T}}{a^{*}} \int_{\Sigma_{A}} \omega_{L}\left(g^{(0)}, \delta g^{(1)}, \mathcal{L}_{\xi_{A}} \delta g^{(1)}\right)+\int_{\Sigma_{A}} \omega_{\phi_{\alpha}}\left(\delta \phi_{\alpha}^{(1)}, \mathcal{L}_{\xi_{A}} \delta \phi_{\alpha}^{(1)}\right).
\end{equation}
Thus, in order to make the cancellation in  Eq.(\ref{eq:second_order_punchline}), we have to specify that 
\begin{equation}
    \frac{\tilde{C}_T}{a^*} = B = \sum_{m=1}^{M} c_m m \frac{(D-2)!}{(D-2m)!} \left(-\frac{1}{\alpha^2}\right)^{m-1} .
    \label{eq:B_constant_constraint}
\end{equation}
This equation will set a constraint on the coefficients $c_m$. Thus, in this sense, we can relax the constraint on the central charges and our results hold regardless of the requirement $\tilde{C}_T/a^* = 1$.
\end{widetext}
\section{Ryu-Takayanagi in Lovelock gravity}
\label{section:ryu-takayanagi_in_lovelock_gravity}
In the AdS/CFT correspondence, a strongly coupled boundary CFT theory corresponds to a weakly coupled string theory in the AdS bulk and the weak-coupling limit of the bulk is classical Einstein gravity. Ryu-Takayanagi\cite{ryu-takayanagi}  proposed that the entanglement entropy for any region is proportional to the area of the minimal surface in the bulk with the same boundary of the region. However, for a general bulk theory, the entanglement entropy formula should be different. The entanglement entropy formula for an arbitrary bulk theory is proposed in \cite{dong2014holographic} and it is shown that a Lovelock gravity bulk has entanglement entropy according to the Jacobson-Meyer formula \cite{jacobson1993black,hung2011holographic}. The entanglement entropy functional for Lovelock is 
\begin{equation}
    S_L = \int_{\Sigma} d^{D-2}y\sqrt{\gamma}\sum_{m=1}^{M(D-2)}m c_m \tilde{L}_{(m-1)}.
    \label{eq:lovelock_entropy}
\end{equation}
Here, $\Sigma$ is a codimension $2$ space-like surface with normal vectors $n_{(\alpha)}$ for $\alpha =1,2$. The surface is shown to be the surface that minimizes the entropy functional. $M(D-2)$ is the largest power of $R$ in dimension $D-2$ space. Also, $\tilde{L}_{(m)}$ is the Lovelock Lagrangian calculated from the intrinsic curvature $\tilde{R}_{ijkl}$, where we used the convention that $i,j,k,l$ mean the indices on $\Sigma$, while $a,b,c,d$ indicate indices in the bulk spacetime. $\gamma_{ij}$ is the induced metric on surface $\Sigma$. 

A closely related entropy formula in Lovelock gravity is Wald entropy formula \cite{Wald_noether}.  Although it is shown in \cite{dong2014holographic}  that in a holographic theory, the entanglement entropy of the boundary theory cannot exactly be calculated by the Wald entropy (in \cite{dong2014holographic} it is shown that the correct EE should be the Jacobson-Myers which differs from Wald's entropy  by precisely the anomaly-like term eq. (3.42) in {\it loc. cit.}), in our argument we use the Wald entropy formula for two reasons. First, as in Ref.\cite{Nonlinear-Faulkner}, we merely show the existence of a bulk gravity theory that satisfies the Lovelock gravity equations to second order and realizes the holographic dual of the boundary CFT. Secondly, we will presently show that all these entropy notions are the same up to second order, at least for ball-shaped regions.  The Wald entropy,  analogous to that of Jacobson-Myers for Lovelock is  given by
\begin{equation}
    S_{LW} = \int_{\Sigma} d^{D-2}y\sqrt{\gamma}\sum_{m=1}^{M(D-2)}m c_m \frak{L}_{(m-1)},
    \label{eq:Wald_Lovelock_entropy}
\end{equation}
where $\frak{L}_{(m-1)}$ is the Lovelock functional for the projected Riemann tensor $\frak{R}_{ijkl}$, and $\Sigma$ is the surface that minimizes the Wald entropy functional. To be specific, the projected Riemann tensor is defined as 
\begin{equation}
    \frak{R}_{ijkl} = \tensor{\gamma}{^a_i} \tensor{\gamma}{^b_j}\tensor{\gamma}{^c_k}\tensor{\gamma}{^d_l} R_{abcd}.
\end{equation}
The projected Ricci tensor is the contraction of the projected Riemann tensor with indices raised by $\gamma^{ij}$ and in the projected Lovelock action, the Kronecker delta is that defined on the surface, and specifically, $\delta^i_j$ on surface is just $\tensor{\gamma}{^i_j}$. 

According to the form of the entropy in Lovelock, it is obvious that the entropy is no longer proportional to the area of the minimal surface regardless of the Wald or Jacobson-Myers constructions and thus we do not expect the RT\cite{RT2} formula to hold anymore. However, we will show in this section that for pure AdS spacetime and for perturbations up to second order, the Wald entropy is the area of the minimal surface. 

\subsection{Pure AdS spacetime}
We first demonstrate the Wald entropy in pure AdS spacetime is proportional to the area of the minimal surface. In pure AdS spacetime, the Riemann curvature tensor is (\ref{eq:ads_curvature}) and therefore, on any surface with projected metric $\gamma_{ij}$, the projected Riemann tensor will be 
\begin{equation}
    \frak{R}_{ijkl} = \tensor{\gamma}{^a_i} \tensor{\gamma}{^b_j}\tensor{\gamma}{^c_k}\tensor{\gamma}{^d_l} R_{abcd} = -\frac{1}{\alpha^2}\left(\gamma_{ik}\gamma_{jl} - \gamma_{il}\gamma_{jk} \right).
    \label{eq:projected_riemann_tensor}
\end{equation}
As a result, the Lagrangian $\frak{L}_{(m-1)}$ will be constant over the surface and therefore, the Wald-entropy functional evaluated over the surface will be 
\begin{equation}
\begin{aligned}
    S_{LW} &= \sum_{m=1}^{M(D-2)} \frac{m c_m(D-2)!}{(D-2m)!}\left(-\frac{1}{{\alpha}^2} \right)^{m-1} \int_{\Sigma} \sqrt{\gamma} d ^{D-2} y \\
    &= C_0 A.
\end{aligned}
\label{eq:entropy_no_perturb}
\end{equation}
Here, we denote the surface as $\Sigma$ and we used the contraction of the Kronecker delta. Also, $A$ denotes the area of the surface $\Sigma$ and $C_0$ is the constant preceeding the integral. Obviously, the Wald entropy formula is proportional to the area of the surface with proportionality constant as $C_0$. Since we did not assume any particular surface in this calculation, the surface that minimizes the functional $S_{LW}$ is the surface with minimal area and thus, we showed that the Wald entropy is proportional to the area of the minimal surface.  
\subsection{First-order perturbation}
Then we focus on the first-order perturbation of the bulk metric. We start with an arbitrary surface in the bulk with the boundary defined by the boundary of the ball-shaped region and when the metric perturbation is performed. Since for small perturbations of the metric, the minimal surface equation produces a surface that is diffeomorphic, in the sense of embedded, to the minimal surface corresponding to vacuum AdS (that is there exists a smooth family of embedding), we can fix coordinates on the surface $\Sigma$ and write the induced perturbed metrics with respect to the fixed coordinates. Therefore, the location of the surface $\Sigma$ is fixed, and the metric is perturbed as $\gamma = \gamma + \epsilon \delta \gamma $. Thus, the  Wald entropy functional will only be varied by changing the projected metric. Note, the functional form of the Wald entropy is similar to the Lovelock gravity action. We can then define 
\begin{equation}
    \frak{P}^{abcd}_{(m)} = \frac{\partial \frak{L}_{(m)}}{\partial \frak{R}_{abcd}}
\end{equation}
as we did in the Lovelock action variation. However, this time, $\delta {\frak{R}_{abcd}} \neq \tensor{\frak{R}}{^{e}_{bcd}} \delta \gamma_{ae} $ and as a result (\ref{eq:variation_of_lovelock_action}) does not hold. Moreover, we notice that $\gamma^{a}_{b}$ is the delta function on the hyperplane and therefore, the variation of it vanishes and as a result, the variation of the projected Riemann tensor will be
\begin{equation}
    \delta \frak{R}_{ijkl} = \tensor{\gamma}{^a_i}\tensor{\gamma}{^b_j}\tensor{\gamma}{^c_k}\tensor{\gamma}{^d_l}\delta R_{abcd} = \tensor{\gamma}{^a_i}\tensor{\gamma}{^b_j}\tensor{\gamma}{^c_k}\tensor{\gamma}{^d_l}\tensor{R}{^e_{bcd}}\delta g_{ae}. 
    \label{eq:projected_riemann_variation}
\end{equation}
Since $\tensor{\gamma}{^{a}_i} g_{ae} = \gamma_{ie}$ and variation of $\tensor{\gamma}{^a_i}$ vanishes, we have $\tensor{\gamma}{^{a}_i}\delta g_{ae} = \delta \gamma_{ie} $. Also, according to the maximally symmetric property of $\tensor{R}{^e_{bcd}}$ which is proportional to $\delta^{e}_c g_{bd} -\delta^{e}_d g_{bc} $, the remaining three projections will recast the Riemann tensor as $\tensor{\frak{R}}{^e_{jkl}}$. As a result, we still have the equality $\delta {\frak{R}_{abcd}} = \tensor{\frak{R}}{^{e}_{bcd}} \delta \gamma_{ae} $ that does not hold in general. Therefore, we can write the variation of the Wald entropy, similar to (\ref{eq:variation_of_lovelock_action}), by defining $\frak{E}^i_j$ analogously as in the Lovelock action for projected Riemann tensors. Also, the functional form of $\frak{E}^i_j$ is also the same as $E^i_j$ by replacing $R_{abcd}$ with $\frak{R}_{ijkl}$. 
Thus, we can write the variation of the Wald functional on the surface as
\begin{equation}
\begin{aligned}
    \delta_1 S_{LW} &= -\int_{\Sigma}d^{D-2} y \sqrt{\gamma} \sum_{m=1}^{M(D-2)} mc_m \frak{E}^{(m-1)}_{ij} \epsilon \delta \tensor{\gamma}{^{ji}} \\
             &= \epsilon \sum_{m=1}^{M(D-2)}m c_m \frac{(D-3)!}{(D-2m-1)!}   \left(-\frac{1}{{\alpha}^2} \right)^{m-1}  \\
             &\qquad \qquad \quad \int_{\Sigma} d^{D-2} y \sqrt{\gamma} \frac{1}{2} \gamma_{ij}\delta \gamma^{ji} = C_1 \delta A.
\end{aligned}
\label{eq:first_order_variation_entropy}
\end{equation}
Therefore, we have shown that the first-order variation of the Wald entropy formula for any surface is proportional to the area but this time, with a different coefficient $C_1$. In order to have a consistent proportional constant, we require $C_0 = C_1$ and this gives a constraint equation as 
\begin{equation}
    \sum_{m=1}^{M(D-2)} \frac{2m-2}{D-2m}  \frac{(D-3)!}{(D-2m-1)!} \left( -\frac{1}{\alpha^2}\right)^{m-1} mc_m = 0.
    \label{eq:entropy_constraint_for_coefficient} 
\end{equation}
Notice, since we did not assume any particular surface $\Sigma$ when performing such a calculation, the surface that minimizes the functional is the surface that minimizes the area. As a result, the Wald surface is the minimal surface in the perturbed spacetime and the Wald entropy is proportional to the area of the surface for first-order perturbations. 

\subsection{Second-order perturbation}
For second-order perturbations, we do not expect an area law from the Wald entropy for Lovelock gravity. However, if we restrict to the fact that the first-order perturbation satisfies the linear Lovelock gravity, we will recover the area law. To show our claim, we will do another variation of the Wald entropy functional, which is just the variation of $\delta_1 S_{LW}$ and we have 
\begin{widetext}
\begin{equation}
    \delta_2 S_{LW} = -\epsilon^2\int_{\Sigma} d^{D-2}y \sqrt{\gamma} \sum_{m=1}^{M(D-2)} mc_m \left(  \sqrt{\gamma} \delta_1 \frak{E}^{k(m-1)}_{\; j}\gamma_{ik} \delta_1\gamma^{ij} + \frak{E}^{k(m-1)}_{\; j}\delta_1 (\sqrt{\gamma}\gamma_{ik} \delta_1 \gamma^{ij})  \right).
    \label{eq:second_order_entropy_variation}
\end{equation}
\end{widetext}
Here the second term is proportional to the variation of the area. The first term can be shown to vanish if we assume first-order Einstein. The first-order Einstein equation shows that \begin{equation}
    \delta R^i_j -\frac{1}{2}\delta^i_j \delta R = 0,
\end{equation}
where $\delta R$ is the trace of $\delta R^i_j$. Thus, if we take the trace, we obtain $\delta R^i_j =0 $. Then, the variation of the projected Ricci tensor will be 
\begin{equation}
    \delta \frak{R}^i_j = \delta \gamma^i_a \gamma^b_j R^a_b +   \gamma^i_a \delta \gamma^b_j R^a_b.
\end{equation}
However, we know that $\delta \gamma^i_a = 0$ and therefore we have $\delta \frak{R}^i_j =0$. As a result, $\delta \frak{R} =0$. Thus, $\delta_1 \frak{E}^{k(m-1)}_{\; j} \sim (\delta \frak{R}^i_j - \frac{1}{2} \delta \frak{R}) =0 $, leading to 
\begin{equation}
\begin{aligned}
    \delta_2 S = \epsilon^2 \sum_{m=1}^{M(D-2)}m c_m \frac{(D-3)!}{(D-2m-1)!}   \left(-\frac{1}{{\alpha}^2} \right)^{m-1} \delta_2  A.
    \label{eq:second_variation_entropy}
\end{aligned}
\end{equation}
This shows that the second variation is also proportional to the area of the surface. Thus, the surface that minimizes the functional will be the surface that minimizes the area and the surface is still the minimal surface. Notice, the proportionality constant is still $C_1$, and we need not introduce new constraints for the Lovelock gravity theory. 

\section{Holographic Entanglement Entropy for Lovelock}\label{entropy}

The discussion in section \ref{section:ryu-takayanagi_in_lovelock_gravity} and in the proof of our main statements only considers the Wald entropy functional for the theory. However, the true holographic entanglement entropy is given by Jacobson-Myers formula evaluated on the surface that minimizes this functional. Thus, in this section, we will show that for spacetimes near the pure AdS spacetime, the holographic entropy for Lovelock gravity is proportional to the area of the surface, and we will show the surface is approximately minimal with error in higher orders. 
If we try to minimize the Jacobson-Myers functional in (\ref{eq:lovelock_entropy}), we will have a constraint equation as 
\begin{equation}
    \sum_m mc_m \tilde{E}^{i(m-1)}_{\;j} K^{j(\alpha)}_{\;i} =0,\quad \alpha =1,2.
    \label{eq:surface_location}
\end{equation}
Here $K^{j(\alpha)}_{\;i}$ is the extrinsic curvature associated with the $\alpha$-th normal vector $n_{(\alpha)}$. $\tilde{E}^{i(m-1)}_{\;j}$ is analogous tensor for $E^i_j$ with $R_{abcd}$ replaced by the instrinsic tensor $\tilde{R}_{abcd}$.  Since the extrinsic curvature is calculated from the embedding function of the surface into the bulk, we will consider the constraint equation as the equation that describes the location of the surface.

\subsection{Pure AdS spacetime}
In pure AdS, we can assume the maximally symmetric form of the Riemann tensor and thus, the projected Riemann tensor is also maximally symmetric on the projection surface as shown in (\ref{eq:projected_riemann_tensor}). According to the Gauss-Codazzi equation,
\begin{equation}
    \frak{R}_{ijkl} = \tilde{R}_{ijkl} + K^{(\alpha)}_{il} K^{(\alpha)}_{jk} - K^{(\alpha)}_{ik} K^{(\alpha)}_{jl},
    \label{eq:gauss-codazzi-equation}
\end{equation}
we are able to associate the projected Riemann tensor with the intrinsic Riemann tensor. As shown in \cite{chen2013note}, the extrinsic curvature tensor $K_{ij}^{(\alpha)}$ vanishes for the minimal surface $\Sigma_B$ with ball-shaped regions as the boundary. Thus, for such minimal surfaces, we have $\frak{R}_{ijkl} = \tilde{R}_{ijkl} $. Also, on that surface, $\tilde{E}^{i(m-1)}_{\;j}\sim \delta^i_j$ and thus (\ref{eq:surface_location}) reduces to $K=0$ which means the location of the surface that minimizes the Jacobson-Myers entropy is the minimal surface. This shows our initial guess is consistent with the constraint. Thus, the minimal surface is the entropic surface. 
\begin{widetext}
\subsection{First-order perturbation}
We consider perturbations, $g = g^{AdS} + \epsilon \delta g$ around pure AdS spacetime. In the perturbed spacetime, assuming we have found the surface $\Sigma'_B$ that minimizes the Jacobson-Myers functional, we expect the extrinsic curvature $K'_{ij}$ on $\Sigma'_B$ will be equal to $K_{ij} + \epsilon \delta K_{ij}$, where $K_{ij}$ is the extrinsic curvature on $\Sigma_B$. Also, we expect the tensor $\tilde{E}_{ij}$ will vary to first-order. Then the constraint equation for the location of the surface will be 
\begin{equation}
    \sum_m mc_m \left\{ \tilde{E}^{i(m-1)}_{\;j} K^{j(\alpha)}_{\;i} + \epsilon \delta \tilde{E}^{i(m-1)}_{\;j} K^{j(\alpha)}_{\;i} + \epsilon \tilde{E}^{i(m-1)}_{\;j} \delta K^{j(\alpha)}_{\;i} + \epsilon^2 \delta \tilde{E}^{i(m-1)}_{\;j}\delta K^{j(\alpha)}_{\;i}  \right\} =0.
    \label{eq:perturbed_surface_location}
\end{equation}
The first two terms are zero due to the fact that on $\Sigma_B$, $K_{ij} =0$ and the last term is of second order. Thus, from the remaining third term and the fact that $\tilde{E}^{i(m-1)}_{\;j} \sim \delta^i_j$, we will have $\delta^i_j\delta K^j_i =0$. Thus, this shows the extrinsic curvature on the new surface $\delta^i_j(K^j_i + \delta K^j_i) =0$. This means that the surface that minimizes the Jacobson-Myers functional is the minimal surface up to second-order in $\epsilon$. In the Holland-Wald gauge, we can fix the location of the surface and thus, the entropy will change to
\begin{equation}
\begin{aligned}
    \delta_1 S_L &= -\int_{\Sigma}d^{D-2} y \sqrt{\gamma} \sum_{m=1}^{M(D-2)} mc_m \tilde{E}^{(m-1)}_{ij} \epsilon \delta \tensor{\gamma}{^{ji}} \\
             &= \sum_{m=1}^{M(D-2)}m c_m \frac{(D-3)!}{(D-2m-1)!}   \left(-\frac{1}{\tilde{\alpha}^2} \right)^{m-1} \epsilon \int_{\Sigma} d^{D-2} y \sqrt{\gamma} \frac{1}{2} \gamma_{ij}\delta \gamma^{ji} = C_1 A.
\end{aligned}
\label{eq:first_order_variation_entropy_jacobson}
\end{equation}
Obviously, we showed $\delta S_L = C_1 \delta A$ with $C_1$ the same as in the Wald entropy case. 
\subsection{Second-order perturbation}
In this section, we assume a Lovelock gravity bulk and for first-order perturbations, we do not have matter fields extrapolated from the boundary CFT and thus, the first-order perturbations satisfy the vacuum Einstein equation, which leads to $\delta R^i_j =0$.  We find again the surfacE $\Sigma'_B$ that minimizes the Jacobson-Myers functional and thus, the extrinsic curvature on the surface is
\begin{equation}
   K''_{ij} = K_{ij} + \epsilon \delta_1 K_{ij} + \delta_2 K_{ij}.  
\end{equation}
Contracting the Gauss-Codazzi equation \ref{eq:gauss-codazzi-equation} leads to
\begin{align}
    &\frak{R}_{ij} = \tilde{R}_{ij} + K^{(\alpha)} K^{(\alpha)}_{ij} - K^{(\alpha)}_{ik} K^{k(\alpha)}_{\;j} \label{eq:codazzi_ricci_tensor} \\
    &\frak{R} = \tilde{R} + K^{(\alpha)} K^{(\alpha)} - K^{(\alpha)}_{ij} K^{ij}_{(\alpha)}.
    \label{eq:codazzi_ricci_scalar}
\end{align}
Since we know $K_{ij} = 0$ when $\epsilon = 0$, the intrinsic curavture $\tilde{R}_{ij}$ and $\tilde{R}$ are equal to their projected counterparts to first-order in $\epsilon$ ( the error is in $O(\epsilon^2)$). Thus, we have 
\begin{equation}
    \delta_1 \tilde{E}^{k(m-1)}_{\; j} \sim \left(\delta_1 \tilde{R}^{i}_j - \frac{1}{2}\delta^i_j \delta_1 \tilde{R}\right) \sim \left(\delta_1 \frak{R}^{i}_j - \frac{1}{2}\delta^i_j \delta_1 \frak{R}\right).
\end{equation}
However, according to the argument in the second-order perturbation analysis in section \ref{section:ryu-takayanagi_in_lovelock_gravity}, we have $\delta_1 \tilde{E}^{k(m-1)}_{\; j} =0 + O(\epsilon^2)$. As a result, the surface location constraint equation for a second-order perturbation of the metric evaluated on the entropic surface will be 
\begin{equation}
    \sum_m mc_m \left\{ \tilde{E}^{i(m-1)}_{\;j} K^{j(\alpha)}_{\;i} + \epsilon^2 \delta_2 \tilde{E}^{i(m-1)}_{\;j} K^{j(\alpha)}_{\;i} + \epsilon^2 \tilde{E}^{i(m-1)}_{\;j} \delta_2 K^{j(\alpha)}_{\;i} + \epsilon^3 \delta_1 \tilde{E}^{i(m-1)}_{\;j}\delta_1 K^{j(\alpha)}_{\;i} \right\} =0,
    \label{eq:perturbed_surface_location_second}
\end{equation}
where we already used the knowledge that $\delta_1 \tilde{E} \sim O(\epsilon^2) $. The first two terms vanish due to the fact that $K_{ij} = 0$ and we neglect higher order perturbations  $O(\epsilon^3)$. Again, we use the fact that $ \tilde{E}^{i(m-1)}_{\;j} \sim \delta^i_j$, we have $\delta^i_j\delta_2 K^j_i =0$. Thus, on the entropic surface, the extrinsic curvature vanishes for perturbations up to second order. Then, the entropic surface is the minimal surface up to second order and we can use the Holland-gauge again to fix the location of the surface and calculate the entropy as 
\begin{equation}
    \delta_2 S_L = -\epsilon^2\int_{\Sigma} d^{D-2}y \sqrt{\gamma} \sum_{m=1}^{M(D-2)} mc_m \left(  \sqrt{\gamma} \delta_1 \tilde{E}^{k(m-1)}_{\; j}\gamma_{ik} \delta_1\gamma^{ij} + \tilde{E}^{k(m-1)}_{\; j}\delta_1 (\sqrt{\gamma}\gamma_{ik} \delta_1 \gamma^{ij})  \right).
    \label{eq:second_order_entropy_variation_jacobson}
\end{equation}
Since the first term vanishes due to the vanishing of $\delta_1\tilde{E}^{k(m-1)}_{\; j}$, we have $\delta_2 S\sim C_1\delta_2 A$. Thus, up to second-order perturbations in Lovelock theory, the holographic entanglement entropy satisfies the area law. 
\end{widetext}

\section{Conclusion} 

Quite generally, we have shown that wedding thermodynamics and entanglement can lead to the more general case of Lovelock rather than the Einstein-Hilbert action.  The intuitive picture behind our construction is that vacuum AdS is symmetric (thus it has constant curvature) and hence is a degenerate spacetime.  This  implies that AdS satisfies both Einstein and (all) Lovelock equations of gravity. Intuitively this suggests that AdS lies at the intersection of many branches of spacetimes representing all the moduli spaces of solutions which are all tangent to one another (first order Einstein and Lovelock coincide) and first order equations cannot suffice to distinguish between all of these branches. Further as we have shown, not even 2nd order equations suffice.  For {\bf generic metrics}, however, we anticipate that  imposing second order conditions suffices to close this loophole.

\section{Acknowledgements} We thank NSF DMR19-19143 for partial funding of this project.

\begin{widetext}
\appendix 
\section{Calculation of $P^{abcd}$ in Lovelock theory}
In this appendix, we will calculate $P^{abcd}$ in Lovelock theory to show explicitly that the symplectic current flow of Lovelock gravity is proportional to that of Einstein theory on the AdS background. We first calculate $P^{abcd}_{(m)}$
\begin{equation}
    \begin{aligned}
    P^{abcd}_{(m)} = \frac{\partial L_{(m)}}{\partial R_{abcd}} = \sum_{i=1}^m \frac{1}{2^m}\delta^{a_1 b_1 \dots a_m b_m}_{c_1 d_1 \dots c_m d_m} \tensor{R}{^{c_1 d_1}_{a_1 b_1}}  \cdots \frac{\partial \tensor{R}{^{c_i d_i}_{a_i b_i}} }{ \partial R_{abcd} }  \cdots \tensor{R}{^{c_m d_m}_{a_m b_m}}.
    \end{aligned}
\end{equation}
Then, we can plug in the form of $\tensor{R}{^{c_j d_j}_{a_j b_j}} $ for $j\neq i$ as (\ref{eq:ads_curvature}) to yield
\begin{equation}
    \tensor{R}{^{c_j d_j}_{a_j b_j}} = -\frac{1}{\alpha^2} \delta^{c_j d_j}_{a_j b_j}.
\end{equation}
Given a quadruple of numbers, $a_i,b_i,c_i,d_i$, it always takes an even number of permutations to interchange $a_ib_i$ and $c_id_i$.   Consequently,  we can we can bring the derivative term to the front and contract the remaining curvature terms  and thus, we have 
\begin{equation}
    P^{abcd}_{(m)} = \frac{m}{2^m} \delta^{a_1 b_1 \dots a_m b_m}_{c_1 d_1 \dots c_m d_m} \left(-\frac{1}{\alpha^2}\right)^{m-1}  \frac{\partial \tensor{R}{^{c_1 d_1}_{a_1 b_1}} }{ \partial R_{abcd} } \delta^{c_2 d_2}_{a_2 b_2}\dots \delta^{c_m d_m}_{a_m b_m}.
\end{equation}
Since $\delta^{c_j d_j}_{a_j b_j}$ is nonvanishing only when $(a_j,b_j) = (c_j,d_j)$ or $(a_j,b_j) = (d_j, c_j)$, the contractions of $\delta^{a_1 b_1 \dots a_m b_m}_{c_1 d_1 \dots c_m d_m}$ with each  $\delta^{c_j d_j}_{a_j b_j}$ would give a factor of $2$ and thus, we find
\begin{equation}
    P^{abcd}_{(m)} = \left(-\frac{1}{\alpha^2}\right)^{m-1} \frac{m}{2} \delta^{a_1 b_1 c_2 d_2\dots c_m d_m}_{c_1 d_1 c_2 d_2 \dots c_m d_m}  \frac{\partial \tensor{R}{^{c_1 d_1}_{a_1 b_1}} }{ \partial R_{abcd} }  = m \frac{(D-2)!}{(D-2m)!} \left(-\frac{1}{\alpha^2}\right)^{m-1} \left( \frac{1}{2} \delta^{a_1b_1}_{c_1d_1} \frac{\partial \tensor{R}{^{c_1 d_1}_{a_1 b_1}} }{ \partial R_{abcd} } \right) .
\end{equation}
The term in the parenthesis is $\partial L_{(1)}/\partial R_{abcd} = P^{abcd}_{(1)}$. Since $L_{(1)}$ is the Einstein theory, we find that on a maximally symmetric spacetime,  the symplectic flow of Lovelock is proportional to that of Einstein. 

\end{widetext}

\clearpage

\end{document}